\documentclass[aps,twocolumn,prl,tightenlines,floatfix,showpacs,superscriptaddress]{revtex4-1}

\usepackage[dvips]{graphicx}
\usepackage[english]{babel}
\usepackage{amsmath}
\usepackage{amssymb}
\usepackage{times}

\begin{document}

\newcommand{\D}{\mathrm{D}}
\newcommand{\p}{\partial}
\newcommand{\Tr}{\mathrm{Tr}}
\renewcommand{\d}{\mathrm{d}}

\title{Phase diagrams of Fermi gases in a trap with mass and population 
imbalances at finite temperature}

\author{Jibiao Wang}

\affiliation{Department of Physics and Zhejiang Institute of Modern
  Physics, Zhejiang University, Hangzhou, Zhejiang 310027, China}
\author{Hao Guo}
\affiliation{Department of Physics, Southeast University, Nanjing 211189, China}
\affiliation{Department of Physics, University of Hong Kong, Hong Kong, China}
\author{Qijin Chen}
\email[Corresponding author: ]{qchen@zju.edu.cn}
\affiliation{Department of Physics and Zhejiang Institute of Modern
  Physics, Zhejiang University, Hangzhou, Zhejiang 310027, China}

\date{\today}

\begin{abstract}
  The pairing and superfluid phenomena in a two-component Fermi gas
  can be strongly affected by the population and mass imbalances. Here
  we present phase diagrams of atomic Fermi gases as they undergo
  BCS--Bose-Einstein condensation (BEC) crossover with population and
  mass imbalances, using a pairing fluctuation theory. We focus on the
  finite temperature and trap effects, with an emphasis on the mixture
  of $^{6}$Li and $^{40}$K atoms. We show that there exist exotic
  types of phase separation in the BEC regime as well as sandwich-like
  shell structures at low temperature with superfluid or pseudogapped
  normal state in the central shell in the BCS and unitary regimes,
  especially when the light species is the majority. Such a
  sandwich-like shell structure appear when the mass imbalance
  increases beyond certain threshold. Our result is relevant to future
  experiments on the $^6$Li--$^{40}$K mixture and possibly other
  Fermi-Fermi mixtures.
\end{abstract}


\maketitle

Ultracold Fermi gases provide an excellent model system for studying
condensed matter physics, owing to the various experimentally tunable
parameters. Using a Feshbach resonance, a two-component Fermi gas of
equal spin mixture of atoms of equal mass exhibits a perfect crossover
from Bardeen-Cooper-Schrieffer (BCS) type of superfluidity to
Bose-Einstein condensation (BEC) \cite{Eagles,Leggett,NSR}, which has
been studied intensively both in experiment
\cite{Ketterle3,Jin4,Thomas2,Salomon3,Grimm4} and theory
\cite{ourreview,GiorginiRMP}. There have also been a great deal of
experimental
\cite{Rice1,*Rice2-old,ZSSK06,*MITPRL06,*Zwierlein2006,*KetterleRF,*MITphase,*Zwierlein2009PRL}
and theoretical \cite{SR06,SM06,Chien06,ChienPRL,YD05,Chevy2010RPP}
studies on equal-mass systems with population imbalance. Population
imbalance adds a new dimension to the phase diagrams, leading to phase
separation \cite{Stability}, Sarma superfluid \cite{Sarma63} and
possibly Fulde-Ferrell-Larkin-Ovchinnikov (FFLO) states \cite{FFLO}.
Mass imbalance, i.e., pairing of different mass atoms, will further
enrich the physics.  Indeed, there have been world wide efforts on the
study of Fermi-Fermi mixture of different species. Over the past
several years, Feshbach resonances between different species of
fermionic atoms, e.g., $^{6}$Li and $^{40}$K, have been found and
studied
\cite{Grimm2008PRL,*Grimm2009PRL,*Grimm2010PRA,*Grimm2011PRL,*Naik,*Grimm2012Nat,Dieckmann2008PRL,*Dieckman2009PRL,*Costa2010PRL,Kokkelmans2010PRL},
although it remains to achieve superfluidity experimentally.  There
have been some theoretical studies in this aspect, e.g., on the strong
attraction limit at zero temperature $T$ \cite{SadeMelo2008PRA},
few-body physics \cite{Blume2012RPP} or the polaron physics
\cite{Stoof2012PRA,Massignan2012EPL} as well as thermodynamics of a
high $T$ normal mixture \cite{Blume2012PRA}.  There have also been
studies on phase diagrams, which, however, are mostly restricted to
zero temperature using a mean field theory either in a homogeneous
Fermi gas \cite{SadeMelo2006PRL,Parish2007PRL} or in a trap
\cite{Duan2006PRA,Yip2006PRB,*Yip2007PRA,Torma2007PRA,guohaoup}. Recently,
Guo \textit{et al.}  \cite{Guo2009PRA} studied the mass imbalanced
Fermi gases at finite temperatures. Due to technical complexity, this
study was restricted to homogeneous systems only. In order to address
various experiments, which are always done at finite $T$ and in a
trap, it is important to take into account the trap and finite
temperature effects simultaneously.

In this paper, we consider a two-species Fermi-Fermi mixture with a
short-range $s$-wave pairing interaction in a 3D isotropic harmonic
trap at finite temperature. We emphasize on the interplay between the
finite $T$ \cite{ourreview,Chen2007PRB} and trap effects while the
mass ratio $m_\uparrow/m_\downarrow$ and population imbalance (or
``spin polarization'') $p = (N_\uparrow - N_\downarrow)/(N_\uparrow +
N_\downarrow)$ (as well as the ratio $\omega_\uparrow
/\omega_\downarrow$ between the trapping frequencies) are varied
within a pairing fluctuation theory, where spin index $\sigma =
\uparrow,\downarrow$ refers to the heavy and light species,
respectively. In order to address experiments, we pay special
attention to the $^{6}$Li--$^{40}$K mixture, while keeping in mind
other possible mixtures such as $^6$Li--$^{173}$Yb and
$^{171}$Yb--$^{173}$Yb. We will present our theoretical findings in
terms of representative phase diagrams in the $T$--$p$ plane, as long
with typical gap and density profiles, throughout the BCS-BEC
crossover.  One special feature here is the emergence of wide-spread
pseudogap phenomena at finite $T$. For high enough mass imbalance, our
result shows that, in the $T$--$p$ phase diagrams, three-shell
sandwich-like spatial structures occupy a large region both at
unitarity and in the BCS regime, including sandwiched phase
separation, sandwiched Sarma, sandwiched polarized pseudogap states
with increasing $T$. In the BEC regime, there are exotic ``inverted''
phase separations with a normal Fermi gas core in the trap center
surrounded by paired (superfluid or pseudogap) state in the outer
shell. Our results provide excellent predictions for future
experiments.

Except for slightly different notations, our formalism is a
combination of that used in Refs.~\cite{Guo2009PRA,ChienPRL}, where a
single-channel Hamiltonian is used to describe the Fermi gas, with a
local density approximation (LDA) for addressing the trap
inhomogeneity.  The heavy ($\sigma = \uparrow$) and light ($\sigma =
\downarrow$) species have dispersion
%
%\begin{eqnarray}
%H&-&\sum_{\sigma}\mu_{\sigma}N_{\sigma}=\sum_{\textbf{k},\sigma}(\epsilon_{\textbf{k},\sigma}-
%\mu_{\sigma})c^{\dag}_{\textbf{k},\sigma}c_{\textbf{k},\sigma}\nonumber\\
%&+&\sum_{\textbf{k},\textbf{k}^{\prime},\textbf{q}}V_{\textbf{k},\textbf{k}^{\prime}}
%c^{\dag}_{\textbf{k}+\textbf{q}/2,\uparrow}c^{\dag}_{-\textbf{k}+\textbf{q}/2,\downarrow}
%c_{-\textbf{k}^{\prime}+\textbf{q}/2,\downarrow}c_{\textbf{k}^{\prime}+\textbf{q}/2,\uparrow},
%\end{eqnarray}
%where $c^{\dag}_{\textbf{k},\sigma}$ creates a particle in the
% momentum state $\textbf{k}$ with spin $\sigma=\uparrow, \downarrow$,
% and
$\xi_{\textbf{k}\sigma}\equiv \epsilon_{\textbf{k}\sigma}-\mu_\sigma
=k^{2}/2m_{\sigma}-\mu_\sigma$, and bare fermion Green's function
$G^{-1}_{0\sigma}(K)=i\omega_{n}-\xi_{\textbf{k}\sigma}$, where
$\mu_{\sigma}$ is the chemical potential, $\omega_n$ the fermionic
Matsubara frequency.  As in Ref.~\cite{Guo2009PRA}, we take
$\hbar=k_{B}=1$ and use the four vector notation, e.g., $K\equiv
(\omega_n, \mathbf{k})$, $\sum_K \equiv T\sum_n\sum_\mathbf{k}$, etc.
At finite $T$, the self-energy $\Sigma_{\sigma}(K)$ contains two
parts,
$\Sigma_{\sigma}(K)=\Sigma_{sc,\sigma}(K)+\Sigma_{pg,\sigma}(K)$,
where the condensate contribution
$\Sigma_{sc,\sigma}(K)=-\Delta_{sc}^{2}G_{0\bar{\sigma}}(-K)$ vanishes
above $T_c$, and the finite momentum pair contribution
$\Sigma_{pg,\sigma}(K)=\sum_{Q}t_{pg}(Q)G_{0\bar{\sigma}}(Q-K)$
persists down to $T=0$. Here $\bar{\sigma} = -\sigma$, and the
$T$-matrix $t_{pg}(Q)=g/[1+g\chi(Q)]$ represents an infinite series of
particle-particle scattering processes, with a short-range interaction
strength $g < 0$ and the pair susceptibility
$\chi(Q)=\sum_{K,\sigma}G_{0\sigma}(Q-K)G_{\bar{\sigma}}(K)/2$. At and
below $T_c$, we have $\Sigma_{pg,\sigma}(K)\approx
\sum_{Q}t_{pg}(Q)G_{0\bar{\sigma}}(-K) =
-\Delta_{pg}^{2}G_{0\bar{\sigma}}(-K) + \delta \Sigma$, which defines
a pseudogap $\Delta_{pg}$ via $\Delta_{pg}^{2}\equiv
-\sum_{Q}t_{pg}(Q)$. Ignoring the less important incoherent term
$\delta\Sigma$, we obtain
$\Sigma_{\sigma}(K)=-\Delta^{2}G_{0\bar{\sigma}}(-K)$ in the simple
BCS form, where
$\Delta^{2}=\Delta_{sc}^{2}+\Delta_{pg}^{2}$. Therefore, the full
Green's function is given by
\begin{equation}
G_{\sigma}(K)=\frac{u_{\textbf{k}}^{2}}{i\omega_{n}-E_{\textbf{k}\sigma}}+
\frac{v_{\textbf{k}}^{2}}{i\omega_{n}+E_{\textbf{k}\bar{\sigma}}},
\end{equation}
where $u_{\textbf{k}}^{2}=(1+\xi_{\textbf{k}}/{E_{\textbf{k}}})/2$,
$v_{\textbf{k}}^{2}=(1-\xi_{\textbf{k}}/{E_{\textbf{k}}})/2$,
$E_{\textbf{k}}=\sqrt{\xi_{\textbf{k}}^{2}+\Delta^{2}}$, and
$E_{\textbf{k}\sigma}=E_{\textbf{k}}+\zeta_{\textbf{k}\sigma}$,
$\xi_{\textbf{k}}=(\xi_{\textbf{k}\uparrow}+\xi_{\textbf{k}\downarrow})/2$,
$\zeta_{\textbf{k}\sigma}=(\xi_{\textbf{k}\sigma}-\xi_{\textbf{k}\bar{\sigma}})/2$. With
$n_{\sigma}=\sum_{K}G_{\sigma}(K)$, $n=n_{\uparrow}+n_{\downarrow}$
and $\delta n=n_{\uparrow}-n_{\downarrow}$, the number equations read
\begin{eqnarray}
  n&=&\sum_{\textbf{k}}\Big\{\Big(1-\frac{\xi_{\textbf{k}}}{E_{\textbf{k}}}\Big)+
  2\bar{f}(E_{\textbf{k}})
  \frac{\xi_{\textbf{k}}}{E_{\textbf{k}}}\Big\},
\label{eq:number}
\\
  \delta
  n&=&\sum_{\textbf{k}}\Big[f(E_{\textbf{k}\uparrow})-f(E_{\textbf{k}\downarrow})\Big],
\label{eq:ndiff}
\end{eqnarray}
where the average Fermi function $\bar{f}(x)\equiv \sum_\sigma
f(x+\zeta_{\textbf{k}\sigma}) /2$.  

At $T\le T_c$ The Thouless
criterion leads to the gap equation $g^{-1}+\chi(0) = 0$.
%
%\begin{equation}
%  \frac{m_{r}}{2\pi a}=\sum_{\textbf{k}}\Big[\frac{1}{2\epsilon_{\textbf{k}}}-\frac{1-
%    2\bar{f}(E_{\textbf{k}})}{2E_{\textbf{k}}}\Big].
%\label{eq:gap1}
%\end{equation}
%
%
For $T>T_{c}$, it is amended by $g^{-1}+\chi(0)=Z\mu_{p}$, where the
effective pair chemical potential $\mu_{p}$ and the coefficient $Z$
can be determined from the Taylor expansion of the inverse $T$-matrix
\cite{ourreview},
$t_{pg}^{-1}(Q)=Z(i\Omega_{l}-\tilde{\Omega}_{\textbf{q}})$, with
$\tilde{\Omega}_{\textbf{q}}=q^{2}/2M^{*}-\mu_p$ being the pair
dispersion, and $M^{*}$ the effective pair mass. Thus the gap equation
reads
\begin{eqnarray}
  \frac{m_{r}}{2\pi a}=\sum_{\textbf{k}}\Big[\frac{1}{2\epsilon_{\textbf{k}}}-
  \frac{1-2\bar{f}(E_{\textbf{k}})}{2E_{\textbf{k}}}\Big]+Z\mu_{p}\,,
\label{eq:gap}
\end{eqnarray}
with $\mu_p = 0$ at $T\le T_c$. Here $g$ is regularized by
$g^{-1}=m_{r}/2\pi a-\sum_{\textbf{k}} 1/2\epsilon^{}_{\textbf{k}}$,
where $a$ is the $s$-wave scattering length,
$m_{r}=m_{\uparrow}m_{\downarrow}/(m_{\uparrow}+m_{\downarrow})$ the
reduced mass, and $\epsilon_{\textbf{k}}=k^{2}/4m_{r} = \xi_\mathbf{k}
+ \mu$, with $\mu = (\mu_\uparrow + \mu_\downarrow)/2$.

The $T$-matrix expansion leads to the pseudogap equation
\begin{equation}
\Delta_{pg}^{2}=Z^{-1}\sum_{\textbf{q}}b(\tilde{\Omega}_{\textbf{q}}),
\label{eq:PG}
\end{equation}
where $b(x)$ is the Bose distribution function.  As in
Ref.~\cite{Guo2009PRA}, we impose a cutoff $q_{c}$ on the summation
such that pairs with $q > q_{c}$ may decay into the particle-particle
continuum.

In a harmonic trap, the LDA approximation imposes that the local
$\mu_{\sigma}(r)=\mu_{\sigma}(0)-\frac{1}{2}m_{\sigma}\omega_{\sigma}^{2}r^{2}$,
with trap frequency $\omega_{\sigma}$. We have
% the overall number constraints:
the total particle number $N=\int \d^{3}r\, n(r)$ and the number
difference $\delta N=N_{\uparrow}-N_{\downarrow}=pN=\int \d^{3}r\,
\delta n(r)$.  The Fermi energy
$E_{F}=(3N)^{1/3}\omega_{\uparrow}=k_{F}^{2}/2m
=mR_{TF}^{2}\omega_{\uparrow}^{2}/2=T_{F}$ is defined as that for an
unpolarized, noninteracting Fermi gas with the same total number $N$
and trap frequency $\omega_{\uparrow}$, with
$m=(m_{\uparrow}+m_{\downarrow})/2$, the average mass. Here $R_{TF}$
% = 2^{1/2}(3N)^{1/6}/(m\omega_\uparrow)^{1/2}$
is the Thomas-Fermi radius, and the species dependent $R_{TF}^\sigma =
\sqrt{2(6N_\sigma)^{1/3}/m_\sigma\omega_\sigma}$.

For a homogeneous system, Eqs. (\ref{eq:number})-(\ref{eq:PG}) form a
closed set and can be used to solve for $T_c$, as well as $\mu$,
$\Delta$, $\Delta_{sc}$ and $\Delta_{pg}$ for given $T$. However, in a
harmonic trap, this set of equations have to be solved at each given
radius and then subject to the total trap-integrated particle number
constraints.

In order to find the stable states in a trap, we compare the (local)
thermodynamical potential $\Omega_{S}$ (per unit volume) in a
superfluid or pseudogap state with its normal Fermi gas counterpart,
$\Omega_{N}=-T\sum_{\textbf{k},\sigma}\ln(1+e^{-\xi_{\textbf{k}\sigma}/T})$. The
paired state $\Omega_{S}$ consists of contributions from fermionic
excitations $\Omega_{F}$ and noncondensed pairs $\Omega_{B}$:
\begin{eqnarray}
  \Omega_{S}&=&\Omega_{F}+\Omega_{B},\\
  \Omega_{F}&=&-\frac{\Delta^{2}}{g}+\sum_{\textbf{k}}(\xi_{\textbf{k}}-E_{\textbf{k}})
  -T\sum_{\textbf{k},\sigma}\ln(1+e^{-E_{\textbf{k}\sigma}/T}),\nonumber\\
  \Omega_{B}&=&Z\mu_{p}\Delta^{2}_{pg}+T\sum_{\textbf{q}}
  \ln(1-e^{-\tilde{\Omega}_{\textbf{q}}/T}).\nonumber
\end{eqnarray}
The stable states should have a lower value. When $\Omega_S >
\Omega_N$ at certain radii $r$, phase separation takes place. Note
that the $\Omega_B$ term is absent in simple mean-field calculations.

\begin{figure}
%\centerline{\includegraphics[clip,width=3.4in] {UnitaryM6.40.eps}}
\centerline{\includegraphics[clip,width=3.4in] {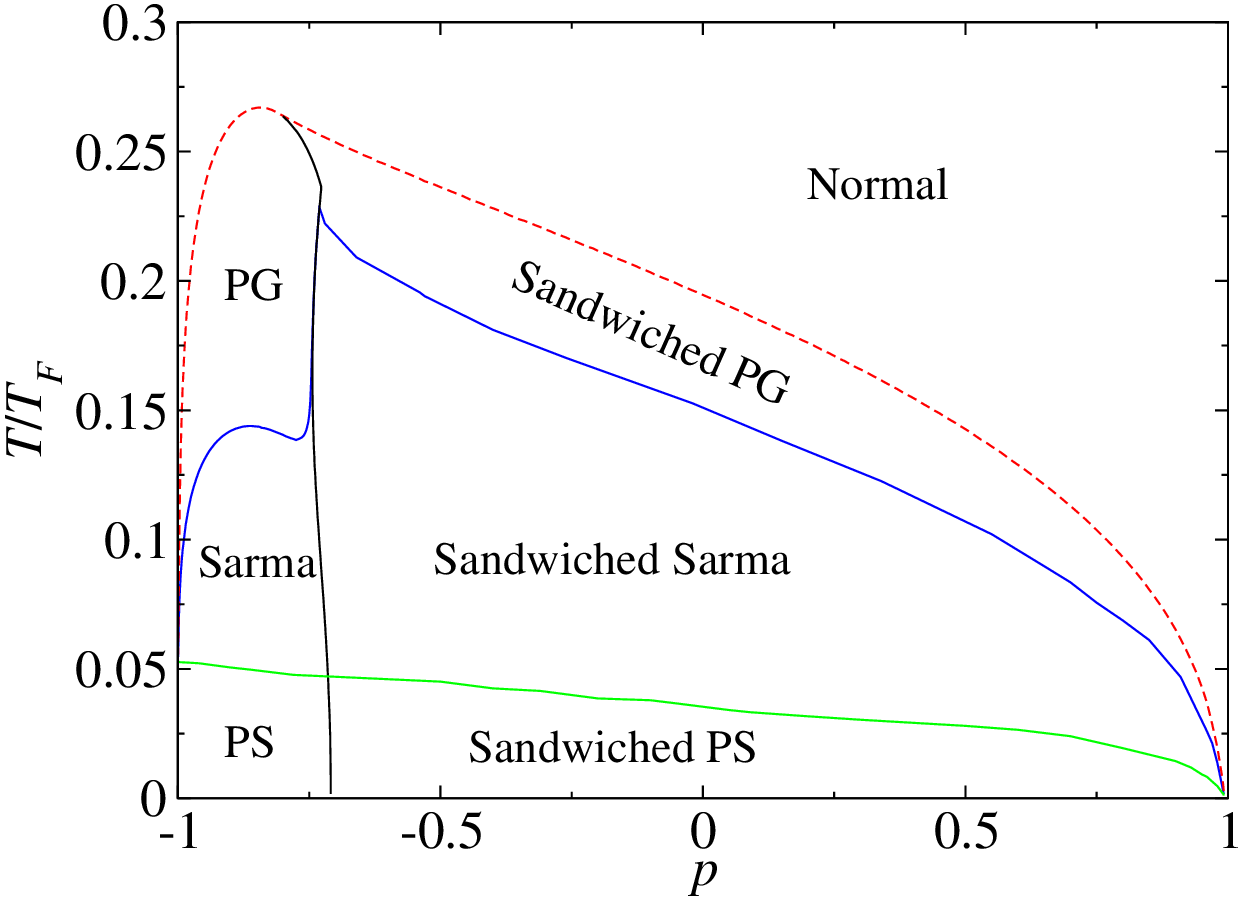}}
\caption{(Color online) $T$--$p$ Phase diagram of $^{6}$Li-$^{40}$K
  mixture in a harmonic trap at unitarity, with
  $\omega_{\uparrow}=\omega_{\downarrow}$.  The solid lines separate
  different phases and the (red) dashed line between the (Sandwiched)
  PG and the normal phases is approximated by mean field
  calculations. Here ``PG" and ``PS" indicate pseudogapped normal
  state and phase separation, respectively. }
\label{fig:Unitary}
\end{figure}

\begin{figure}
\centerline{\includegraphics[clip,width=3.4in] {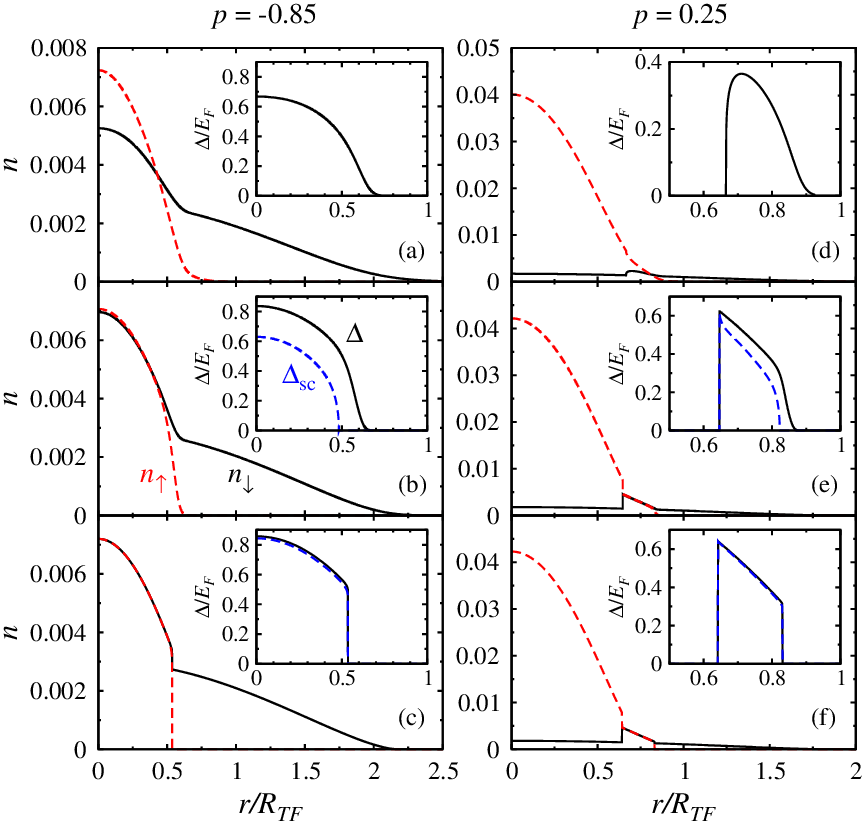}}
%\centerline{\includegraphics[clip,width=3.4in] {Unitary-gap-density.eps}}
\caption{(Color online) Representative density (main panels) and gap
  (insets) profiles for $p= -0.85$ and 0.25 at unitarity: (a)-(c)
  $T/T_F=0.2$, $0.1$, $0.02$; (d)-(f) $T/T_F=0.15$, $0.05$, $0.01$,
  corresponding to each phase in Fig.~\ref{fig:Unitary}. The
  temperatures in (c) and (f) are essentially zero. The (black) solid
  and (red) dashed lines in the main panels are for $^{6}$Li
  ($n_{\downarrow}$, light species) and $^{40}$K ($n_{\uparrow}$,
  heavy species), respectively. Plotted in the insets are the total
  gap $\Delta$ (black solid lines) and order parameter $\Delta_{sc}$
  (blue dashed lines).  Here the insets share the same x-axes, and 
% $R_{TF}=\sqrt{4(3N)^{1/3}/[(m_{\uparrow}+m_{\downarrow})\omega_{\uparrow}]}$  is the Thomas-Fermi radius and 
%  the number 
  the densities $n_{\sigma}$ are in units of $n_{F}^{}\equiv
  k^{3}_{F}/3\pi^{2}$. }
\label{fig:Ugd}
\end{figure}

Shown in Fig.~\ref{fig:Unitary} is the calculated $T$--$p$ phase
diagram at unitarity $1/k_{F}a=0$ with
$\omega_{\uparrow}=\omega_{\downarrow}$ and mass ratio $m_\uparrow
/m_\downarrow =$ 40:6, as is appropriate for the $^{6}$Li-$^{40}$K
mixture. Corresponding representative density and gap profiles are
shown in Fig.~\ref{fig:Ugd}. It can be seen that phase separation (PS)
and sandwiched PS occupy the lowest $T$ part. In particular, when the
light species dominates the population, we have a regular phase
separation with an equal population superfluid core in the trap
center, surrounded by the majority light atoms in the outer shell,
similar to that seen in the equal-mass case \cite{ChienPRL}.  Here the
gap $\Delta$, the order parameter $\Delta_{sc}$, and the density of
minority atoms $n_\downarrow$ jump to zero across the interface, as
shown in Fig.~\ref{fig:Ugd}(c). However, except for the light atom
dominated case, a three shell structure appears, which we refer to as
sandwiched PS. Typical density and gap profiles are shown in
Fig.~\ref{fig:Ugd}(f) (for essentially zero $T$), where an equal
population superfluid exists only at intermediate radii, whereas the
light and heavy atoms dominate the outer and inner shells,
respectively.  In fact, at zero $T$, the heavy atoms are absent in the
outer shell. The gaps and density profiles exhibit first order jumps
at both interfaces. Our findings at the lowest $T$ are consistent with
the earlier works at zero temperature \cite{Duan2006PRA,guohaoup}. The
primary reason for this three shell structure to occur is that
$R_{TF}^\uparrow \ll R_{TF}^\downarrow$ so that $n_\uparrow \gg
n_\downarrow$ at the trap center. It is known that at low $T$,
population imbalance tends to break pairing \cite{Chien06}. Therefore,
only at certain intermediate radii where $n_\uparrow \approx
n_\downarrow$ 
% the densities of the two species are close to each other
can a superfluid exist. As $p$ decreases from 1 towards -1, these
radii move from the trap edge near $R_{TF}^\uparrow$ towards the trap
center, until the inner shell of normal mixture
% of heavy and light atoms
shrinks to zero around $p = -0.71$.

Previous study \cite{Chien06} demonstrates that in a homogeneous Fermi
gas of equal mass at unitarity, a population imbalanced Sarma
superfluid \cite{Sarma63} is stable only at intermediate
temperatures. Here we find this remains true for an unequal mass
mixture. In Fig.~\ref{fig:Unitary}, the PS phase becomes a Sarma phase
at intermediate $T$, where population imbalance penetrates into the
inner superfluid core so that the first order jumps of the gap, the
order parameter and the minority density at the interface
disappear. This is similar to the Sarma state found in
Ref.~\cite{ChienPRL}. Correspondingly, the sandwiched PS phase becomes
sandwiched Sarma phase. As seen from the $p=-0.85$ and 0.25 cases
shown in Figs.~\ref{fig:Ugd}(b) and \ref{fig:Ugd}(e), respectively,
$\Delta$, $\Delta_{sc}$ and $n_\uparrow$ vanish continuously at the
interface (of the larger radius). The difference between $\Delta$ and
$\Delta_{sc}$ defines the presence of the pseudogap (not shown)
$\Delta_{pg} = \sqrt{\Delta^2 - \Delta_{sc}^2}$.

As the temperature increases further, the superfluid region disappears
so that the Sarma phase becomes a polarized pseudogap (PG) phase,
where a finite pseudogap exist in the inner core without
superfluidity. In a similar fashion, the sandwiched Sarma phase
evolves into a sandwiched PG phase. Representative density and gap
profiles are shown in Figs.~\ref{fig:Ugd}(a) and \ref{fig:Ugd}(d) for
($p$, $T/T_F$) = (-0.85, 0.2) and (0.25, 0.15), respectively.
% Note that the blue solid line which separates the (sandwiched) PG
% and Sarma phases in Fig.~\ref{fig:Unitary} is continuous between $p
% = \pm 1$.
Finally, at very high $T$ we have a normal phase. Note that the
separation between the normal and the PG or sandwiched PG phases is a
crossover rather than a phase transition, and thus is determined
approximately using the BCS mean-field theory.

\begin{figure}
%\centerline{\includegraphics[clip,width=3.0in] {BCS_BEC_M6.40.eps}}
\centerline{\includegraphics[clip,width=3.0in] {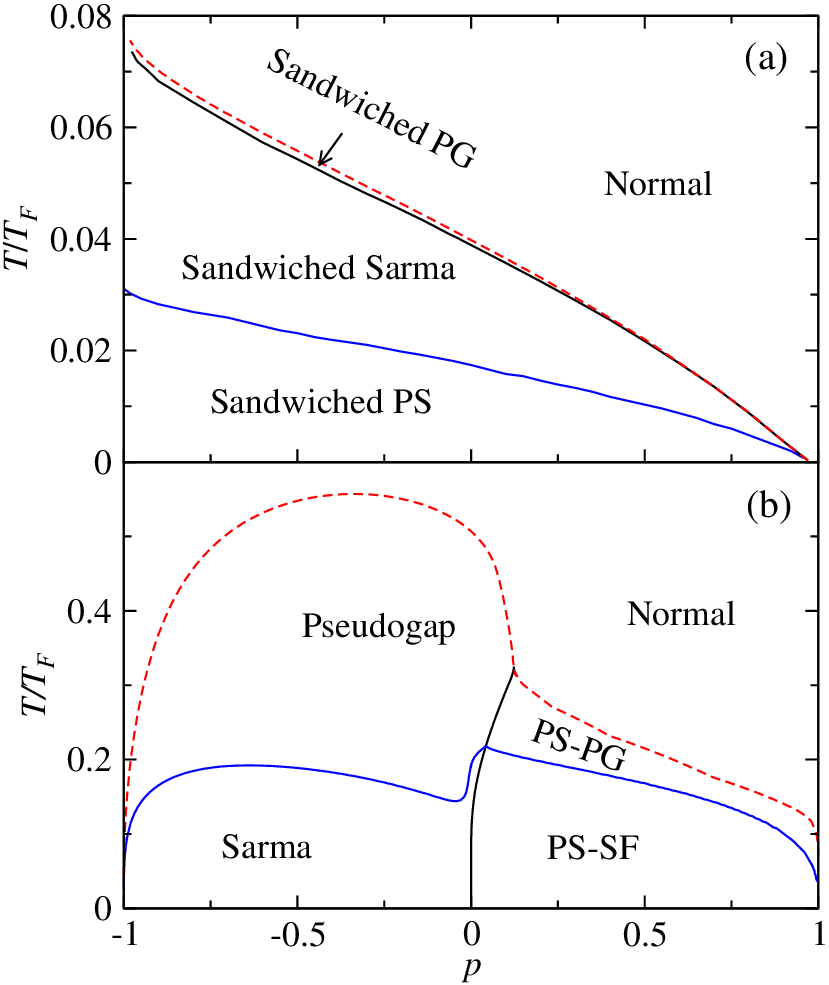}}
\caption{(Color online) Phase diagram of $^{6}$Li-$^{40}$K as in
  Fig.~\ref{fig:Unitary} but for (a) $1/k_{F}a=-0.5$ and (b)
  +0.5. Here ``PS-SF'' and ``PS-PG'' represent phase separated
  superfluid and phase separated pseudogap phases, respectively, with
  a normal gas core surrounded by an outer shell of superfluid or
  pseudogapped normal mixture.  }
\label{fig:BCS_BEC}
\end{figure}

\begin{figure}
%\centerline{\includegraphics[clip,width=3.4in] {BCS-BEC-gap-density.eps}}
\centerline{\includegraphics[clip,width=3.4in] {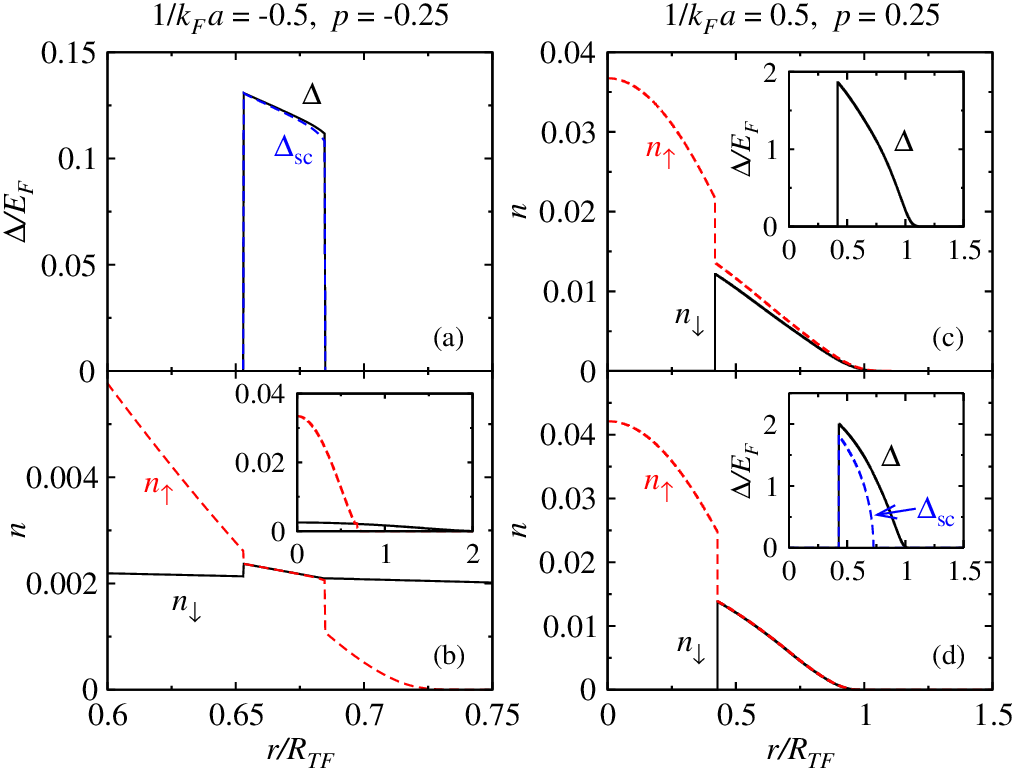}}
\caption{(Color online) Typical density and gap profiles for BCS (left
  column, $1/k_Fa = -0.5$) and BEC (right column, $1/k_Fa = 0.5$)
  regimes. The convention is the same as in Fig.~\ref{fig:Ugd}. Panels
  (a) and (b) are for $p=-0.25$ at $T=0.01T_{F}\approx 0$,
  corresponding to the sandwiched PS phase in
  Fig.~\ref{fig:BCS_BEC}(a).  Panels (c) and (d) plot the density and
  gap (insets) distributions for $p = 0.25$ at $T=0.2T_{F}$ and
  $0.1T_{F}$, representing the PS-PG and PS-SF phases in
  Fig.~\ref{fig:BCS_BEC}(b), respectively. }
\label{fig:BCS-BECgd}
\end{figure}

Next, we present in Fig.~\ref{fig:BCS_BEC} the phase diagrams at (a)
1/$k_{F}a$=-0.5 and (b) 0.5, similar to Fig.~\ref{fig:Unitary}, but in
the (near-)BCS and (near-)BEC regimes, respectively.  For the BCS
case, except for the high $T$ normal phase, the phase diagram is
essentially occupied by three-shell sandwich-like structures. The
central shell is an unpolarized BCS superfluid at the lowest $T$, spin
polarized Sarma superfluid at intermediate $T$, and a polarized
pseudogapped normal state at slightly higher $T$. In the sandwiched PS
phase, the outer shell is a normal mixture with light species
($^{6}$Li) in excess, surrounded by a single component normal gas of
the light atoms at the trap edge. This is different from the
sandwiched PS at unitarity [Fig.~\ref{fig:Ugd}(f)], where the outer
shell contains the light atoms only at $T=0$. Typical gap and density
profiles (at $p = -0.25$ near $T=0$) are shown in
Figs.~\ref{fig:BCS-BECgd}(a) and \ref{fig:BCS-BECgd}(b), respectively.
In comparison with the unitary case, one can see that as the pairing
strength decreases from unitarity, the sandwiched PS phase in the
phase diagram expands and gradually squeezes out the PS phase
completely. The evolution of the various phases with increasing
temperature is similar to their unitary counterparts, except that now
the sandwiched PG phase occupies a very slim region, reflecting a much
weaker pseudogap effect in the BCS regime.

The phase diagram for the BEC case in Fig.~\ref{fig:BCS_BEC}(b) is
rather different. First, for $p<0$, where the light species is the
majority, a Sarma superfluid phase occurs at low $T$, with essentially
equal population at the trap center. Indeed, polarized superfluid
becomes stable and phase separation is no longer the ground state in
the BEC regime \cite{Chien06,Stability}. As $T$ increases, the order
parameter decreases to zero and the system evolves into a polarized
pseudogapped normal state. The large area of the ``pseudogap'' phase
indicates greatly enhanced pseudogap effects in the BEC regime. On the
other hand, for (roughly) $p>0$, where the heavy species dominates, we
have an ``inverted'' phase separated superfluid state at low $T$,
labeled as ``PS-SF'', where a normal gas core of the heavy species is
surrounded by a shell of unpolarized superfluid. This should be
contrasted with the PS phase in the unitary case
[Fig.~\ref{fig:Ugd}(c)], where the normal Fermi gas is outside the
superfluid core. As $T$ increases, a phase separated pseudogap state
(labeled ``PS-PG'') appears, where pseudogap exists in the polarized
outer shell but without superfluidity. This is an exotic new phase,
which has never been seen or predicted before. Typical density and gap
(insets) profiles for the PS-PG and PS-SF phases are shown in
Figs.~\ref{fig:BCS-BECgd}(c) and \ref{fig:BCS-BECgd}(d), respectively.
Possible causes for the ``inversion'' of the phase separation include:
(i) For $p>0$, $R_{TF}^\uparrow$ of the heavy species becomes close to
$R_{TF}^\downarrow$ of the light atoms; (2) As the local $k_F(r)$
decreases with $r$, the outer region is deeper in the BEC regime than
the trap center, making pairing easier and energetically more
favorable at the trap edge. When compared with the three-shell
structure at unitarity, one concludes that as the pairing strength
increases from unitarity, the outer shell of normal light atoms
retreats and finally disappears.

\begin{figure}
%\centerline{\includegraphics[clip,width=3.in] {UnitarityP-0.23_mr_wr.eps}}
\centerline{\includegraphics[clip,width=3.in] {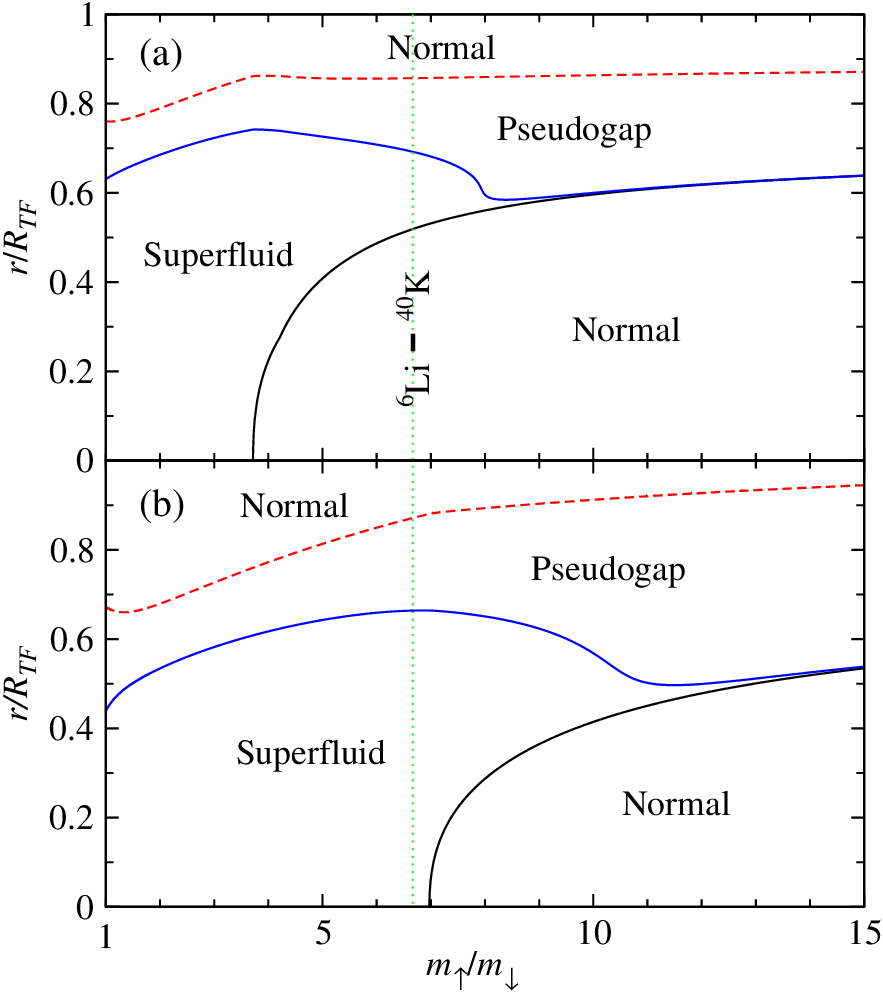}}
\caption{(Color online) Spatial distributions of various phases in the
  trap at unitarity as a function of mass ratio
  $m_{\uparrow}/m_{\downarrow}$ for $p=-0.23$ with (a) $T=0.1T_{F}$,
  $\omega_{\uparrow}=\omega_{\downarrow}$ and (b) $T=0.15T_{F}$ and
  $\omega_{\downarrow}=2\omega_{\uparrow}$. The (green) dotted line
  indicates the mass ratio for $^6$Li-$^{40}$K.}
\label{fig:mratio}
\end{figure}

%\begin{figure}
%\centerline{\includegraphics[clip,width=3.4in] {UnitarityP-0.23T0.1+0.15_mr_wr_2.eps}}
%\caption{(Color online) Spatial distributions of various phases in the
%  trap at unitarity as a function of mass ratio
%  $m_{\uparrow}/m_{\downarrow}$ for $p=-0.23$ with (a) $T=0.1T_{F}$,
%  $\omega_{\uparrow}=\omega_{\downarrow}$ and (b) $T=0.15T_{F}$ and
%  $\omega_{\downarrow}=2\omega_{\uparrow}$. The (green) dotted line
%  indicates the mass ratio for $^6$Li-$^{40}$K.}
%\label{fig:mratio}
%\end{figure}

We now turn to the case of variable mass ratio
$m_{\uparrow}/m_{\downarrow}$, with different
$\omega_\uparrow/\omega_\downarrow$. Plotted in Fig.~\ref{fig:mratio}
are spatial distributions of possible phases at unitarity as a
function of the mass ratio at $p=-0.23$ for (a) $(T/T_F,
\omega_\uparrow/\omega_\downarrow) = (0.1, 1)$ and (b) (0.15, 1/2),
respectively. For both cases, a sandwich-like structure appears as the
mass ratio increases beyond 3.7 and 7.0, respectively. This can be
easily understood by looking at the corresponding non-interacting
density distributions \cite{Duan2006PRA}.  Starting from
$m_{\uparrow}/m_{\downarrow} = 1$ and
$\omega_\uparrow/\omega_\downarrow = 1$, the majority species always
has a larger spatial extension so that pairing is easier at the trap
center. However, as $m_{\uparrow}/m_{\downarrow}$ becomes sufficiently
large, $R_{TF}^\uparrow$ of the heavy species may become smaller than
$R_{TF}^\downarrow$ so that the non-interacting density distribution
curves cross each other at an intermediate radius, where pairing is
more energetically favorable than elsewhere so that a three-shell
structure appears at low $T$, as shown in
Fig.~\ref{fig:mratio}(a). Note that BCS pairing requires a match of
(mass independent) $k_{F}^{\sigma} = (6\pi^2 n_\sigma)^{1/3}$
(locally) between the two species. Thus the mass ratio changes the
position of the crossing point via changing $R_{TF}^\sigma$.

Tuning $\omega_\uparrow/\omega_\downarrow$ can also change the density
crossing point since $R_{TF}^\sigma$ depends on the product
$m_\sigma\omega_\sigma$, as shown in Fig.~\ref{fig:mratio}(b).  This
explains why the threshold $m_{\uparrow}/m_{\downarrow}$ for the
three-shell structure to occur in Fig.~\ref{fig:mratio}(b) is roughly
twice that in Fig.~\ref{fig:mratio}(a).  The (green) vertical dotted
line indicates where the $^6$Li--$^{40}$K mixture resides. In this
case, a three-shell structure appears in the upper panel while only a
regular Sarma phase shows up in the lower panel of
Fig.~\ref{fig:mratio}.

It is worth pointing out that at large mass imbalance, the strong
disparity between $R_{TF}^\uparrow$ and $R_{TF}^\downarrow$ makes
population balance or imbalance less important.

We end by noting that we have not included the species dependent,
incoherent part of the fermion self energy, which may induce polarons
in the mixed normal states, e.g., in the inner core of the three-shell
structured phases. However, it is not important for the present study
whether or not the minority species in these states form polarons.
Following common practice
\cite{Kadanoff,Leggett,NSR,Randeria2,*Randeria1997,Chen2,Strinati4,Griffin5,Drummond2,Tchern,Haussmann2},
we have also neglected the particle-hole channel contributions
\cite{GMB}, which can be roughly approximated by a shift in the
pairing interaction strength \cite{Yin2009,ParticleHoleChannel}. These
approximations are expected to modify the phase boundaries \emph{only
  quantitatively}. In addition, we have not considered the FFLO states
which, in an equal-mass Fermi gas, appears to be of less interest in 
3D 
% at unitarity and in the BEC regime but only at very low temperature
% and in a small region on the BCS side
\cite{LOFF1,SR06,Duan2007PRA}.  Effects of mass imbalance on the FFLO
phases will be investigated in a future work.

In summary, we have studied the finite temperature phase diagrams for
Fermi gases in a trap with both mass and population imbalances, using
a pairing fluctuation theory, with special attention paid to the
$^6$Li--$^{40}$K mixture. Unique to our theory are the wide spread
pseudogap phenomena and the prediction of exotic phases, e.g., the
phase-separated pseudogap phase, which can be tested by measuring the
density and gap profiles in the trap. In particular, vortex
measurements and rf spectroscopy \cite{Chen_RPP} may
be used to ascertain the superfluid and pseudogapped normal states. In
order to compare with concrete experiments, detailed parameters such
as $N_\sigma$, $\omega_\sigma$, etc are needed. Our results can be
tested experimentally when such experiments become available in the
(near) future.

This work is supported by NSF of China (Grant No. 10974173), the
National Basic Research Program of China (Grants No. 2011CB921303 and
No. 2012CB927404), and Fundamental Research Funds for Central
Universities of China (Program No. 2010QNA3026).

%\vspace*{-1ex}

\bibliographystyle{apsrev} 
%\bibliography{Review3}

\end{document}